\documentclass[journal]{IEEEtran}
\usepackage{cite}

\ifCLASSINFOpdf
 \usepackage[pdftex]{graphicx,color,psfrag}  
\else
\usepackage[dvips]{epsfig,graphicx,color,psfrag}
\fi
\begin{document}
%
\title{A Reliability of Measurement Based Algorithm for Adaptive Estimation in Sensor Networks}
%
%
%

\author{Wael M. Bazzi,
        Amir~Rastegarnia
        ~and~Azam~Khalili 
\thanks{W. M. Bazzi is with Electrical and Computer Engineering Department, American University in Dubai, Dubai, United Arab Emirates, email: wbazzi@aud.edu}
\thanks{A. Rastegarnia, and A. Khalili are with  Department of Electrical Engineering, Malayer University, Malayer 65719-95863, Iran e-mails: (rastegar,tinati,a-khalili@tabrizu.ac.ir).}}

\maketitle

\begin{abstract}
In this paper we consider the issue of reliability of measurements in distributed adaptive estimation problem. To this aim, we assume a sensor network with different observation noise variance among the sensors and propose new estimation method based on incremental distributed least mean-square (IDLMS) algorithm. The proposed method contains two phases: I) Estimation of each sensor's observation noise variance, and
II) Estimation of the desired parameter using the estimated observation variances. To deal with the reliability of measurements, in the second phase of the proposed algorithm, the step-size parameter is adjusted for each sensor according to its observation noise variance. As our simulation results show, the proposed algorithm considerably improves the performance of the IDLMS algorithm in the same condition.
\end{abstract}


\IEEEpeerreviewmaketitle

\section{Introduction}
%
%
%
%
\IEEEPARstart{C}{onsider} a wireless sensor network composed of distributed sensor nodes as shown in Fig. 1. The purpose is to  estimate an unknown vector $w^o$ from multiple spatially independent but possibly time-correlated measurements collected at $N$ nodes in a network. Each node $k$ has access to time-realizations $\left\{d_k(i),u_{k,i}\right\}$  of zero-mean spatial data $\left\{d_k,u_k\right\}$ where each $d_k$ is a scalar measurement and each $u_k$ is a $1\times M$ row regression vector. We assume that the unknown vector  relates to the  as:
\begin {equation}
d_k(i)=u_{k,i}w^o+v_k(i),
\end {equation}
where $v_k(i)$ is observation noise with variance $\sigma^{2}_{v,k}$ and is independent of $\left\{d_k(i),u_{k,i}\right\}$. A number of studies have considered such a distributed estimation problem \cite{est01,xiao06,sayed14a}. In \cite{lop06,sayed06,lop07,li10,lop10,khalili10,cat11,khalili12,rast14} distributed adaptive estimation algorithms using incremental optimization techniques are developed and their transient and steady-state performance analysis are also provided. The IDLMS and distributed recursive least mean-square (DRLS) \cite{sayed06} are the examples of such algorithms. These algorithms are distributed, cooperative, and able to respond in real time to changes in the environment. In these algorithms, each node is allowed to communicate with its immediate neighbor in order to exploit the spatial dimension while limiting the communications burden at the same time. In \cite{lop08,cat08,cat09,tak09,cat10,gha13,di14}, diffusion implementation of distributed adaptive estimation algorithms are developed. In these algorithms, each node can communicate with all its neighbors as dictated by the network topology. Both LMS-based and RLS-based diffusion algorithms are given in the literature. In addition, for both of these cases the performance analysis can be found in \cite{lop07} and \cite{lop08} respectively. In comparison with \emph{incremental} based algorithms, \emph{diffusion} based methods need more communication resources while have better estimation performance. Both diffusion LMS and diffusion RLS algorithm are introduced in the literature. 

In all of the mentioned distributed adaptive estimation algorithms, either equal observation noise is assumed for all the nodes in the network or same strategy is used for different variance condition. The motivation for a new estimation method stems from the following facts: 1) The equal observation noise variance is not a suitable assumption in practice, and 2) It is clear that if the issue of reliability of observations is considered, better estimation performance can be expected. In this paper, to deal with the mentioned problems and especially the issue of reliability of observations, we propose a new distributed adaptive estimation algorithm. In the proposed method which is based on IDLMS, first each sensor's observation noise variance is estimated and in the next step, based on the estimated variances, the step-size parameter is adjusted according to estimated observation noise variances.

\begin{figure}[t]
\centering 
\includegraphics [width=4cm] {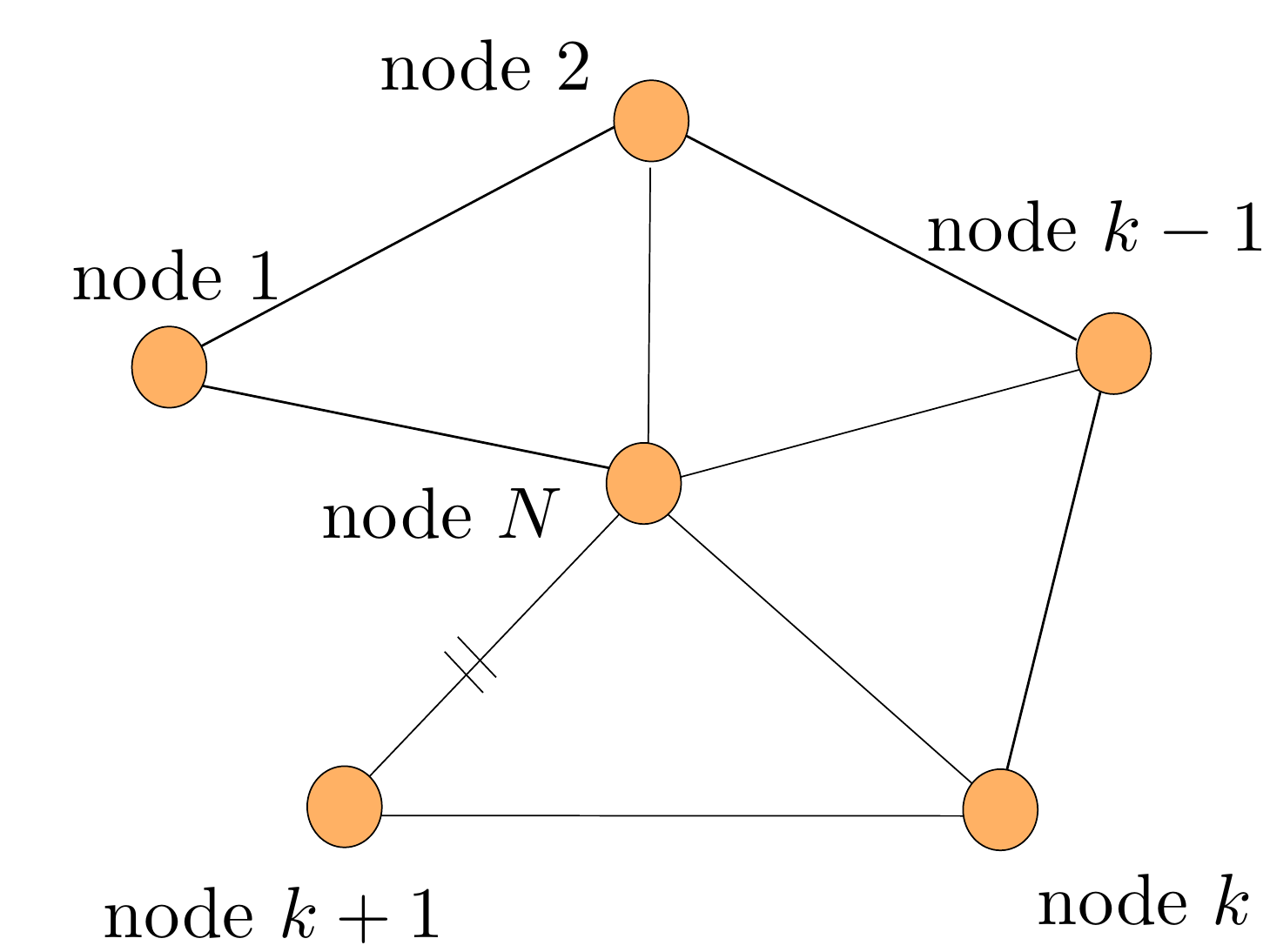}   
\centering\caption{A distributed network with $N$ sensor nodes.}
\end{figure}

\section{Estimation Problem And The Adaptive Distributed Solution}
\subsection{Notation and Assumptions}
A list of the symbols used through the paper, for ease of reference, are shown in Table I.

\begin{table}[h]
\renewcommand{\arraystretch}{2}
\caption{List of the Main Used Symbols}
\centering
\begin{tabular}{cl}
\hline \hline
\ \ Symbol\ \ \ \ & \ \ \ \ {Description} \ \ \\
\hline
\ \ $w_i$ \ \ \ \ &\ \ \ \ Weight vector estimate at iteration $i$\ \ \\
\ \ $u_i$ \ \ \ \ &\ \ \ \ Regressor vector at iteration $i$\ \ \\
\ \ $e(i)$ \ \ \ \ &\ \ \ \ Output estimation at iteration $i$\ \ \\
\ \ $d(i)$ \ \ \ \ &\ \ \ \ Value of a \emph{scalar} variable $d$ at iteration $i$\ \ \\
\ \ $u_i$ \ \ \ \ &\ \ \ \ Value of a \emph{vector} variable $u$ at iteration $i$\ \ \\
\hline \hline
\end{tabular}
\end{table}

The subsequent equations rely on the following assumptions
\begin{itemize}
	\item $u_{k,i}$ is independent of $u_{l,i}$ for $k \ne l$, (spatial independence).
	\item For every $k$, the sequence $u_{k,i}$ is independent over time (time independence).
	\item The variances of observation noise for all of the sensors do not vary with time.
\end{itemize}

\subsection{problem Statement}
By collecting regression and measurement data into global matrices results (see 1):
\begin{equation}
U\buildrel \Delta \over = col\left\{u_1,u_2,...,u_N\right\}
\end{equation}
\begin{equation}
d\buildrel \Delta \over = col \left\{d_1,d_2,...,d_N\right\}
\end{equation}
where the notation $\rm{col}\left\{\cdot\right\}$ denotes a column vector (or matrix) with the specified entries stacked on top of each other. The objective is to estimate the  vector  that solves
\begin{equation}
\min_{w} J(w)\ \textsl{where}\ J(w)=E\left\|d-Uw\right\|^2
\end{equation}
The optimal solution  satisfies the normal equations \cite{lop07}
\begin{equation}
R_{du}=R_uw^o
\end{equation}
where
\begin{equation}
R_{du}  = E\left\{ {U^* d} \right\} = \sum\limits_{k = 1}^N {R_{du,k} } ,
\end{equation}
\begin{equation}
R_{uu}  = E\left\{ {U^* U} \right\} = \sum\limits_{k = 1}^N {R_{u,k} } .
\end{equation}
where in (6), the symbol * denotes the Hermitian
transform. Note that in order to use (5) to compute
$w^o$ each node must have access to the global statistical
information $\left\{R_u,R_{du}\right\}$ which in turn requires more
communications between nodes and computational
resources.

\subsection{Incremental LMS solution}
The standard gradient-descent implementation to solve the normal equation (5) is as
\begin{equation}
w_i  = w_{i - 1}  + \mu \left[ {\nabla J\left( {w_{i - 1} } \right)} \right]^ *  ,
\end{equation}
where $\mu$  is a suitably chosen step-size parameter, $w_i$ is an estimate for desired parameter (i.e. $w_o$) in $i$th iteration of $\nabla J( \cdot )$ and denotes the gradient vector of $J(w)$ evaluated at $w_{i - 1} $ . If $\mu$  is sufficiently small then $w_i  \to w^o $ as $i \to \infty $ \cite{lop06,sayed06,lop07}. In order to obtain a distributed version of (8), first the cost function $J(w)$  is decomposed as
\begin{equation}
J(w) = \sum\limits_{k = 1}^N {J_k (w)},
\end{equation}
where
\begin{equation}
J_k  \buildrel \Delta \over = E\left\{ {\left| {d_k  - U_k w} \right|^2 } \right\}.
\end{equation}
Using (9) and (10) the standard gradient-descent implementation of (8) can be rewritten as [3-6]:
\begin{equation}
w_i  = w_{i - 1}  - \mu \left[ {\sum\limits_{k = 1}^N {\nabla J_k \left( {w_{i - 1} } \right)} } \right]^ *
\end{equation}
By defining the  as the local estimate of the $\psi _k^{(i)} $ at node $k$ and time $i$, then $w_i $ can be evaluated as 
\begin{equation}
\psi _k^{(i)}  = \psi _{k - 1}^{(i)}  - \mu \left[ {\nabla J_k \left( {w_{i - 1} } \right)} \right]^ *  ,\;k = 1,2, \ldots ,N
\end{equation}
This scheme still requires all node to share global information $w_{i - 1} $. The fully distributed solution can be achieved by using the local estimate $\psi _k^{(i)} $ at each node $k$ instead of $w_{i - 1} $,
\begin{equation}
\psi _k^{(i)}  = \psi _{k - 1}^{(i)}  - \mu \left[ {\nabla J_k \left( {\psi _k^{(i)} } \right)} \right]^ *  ,\;k = 1,2, \ldots ,N
\end{equation}
Now, we need to determine the gradient of $J$ and replace it in (13). To do this, the following approximations are used
\begin{equation}
R_{du,k}  \approx d_k (i)\,u_{k,i}^ *  
\end{equation}
\begin{equation}
R_{u,k}  \approx u_{k,i}^ *  u_{k,i}
\end{equation}
The resulting IDLMS algorithm is as follows 
\begin{equation}
\left\{ \begin{array}{l}
 \psi _0^{(i)}  \leftarrow w_{i - 1}  \\ 
 \psi _k^{(i)}  = \mathop {\psi _{k - 1}^{(i)}  - \mu }\limits_{k = 1,2, \ldots ,N} u_{k,i}^* \left[ {d_k (i) - u_{k,i} \psi _{k - 1}^{(i)} } \right]\, \\ 
 w_i  \leftarrow \psi _N^{(i)}  \\ 
 \end{array} \right.
\end{equation}

\section{Proposed Algorithm}
\subsection{Motivation}
As mentioned in the introduction section, equal observation noise assumption for all nodes could not comply with situations in physical problems. On the other hand, although considering some noisy sensors in the network (as in \cite{rast08a}) is a better assumption for sensor network, but it is still far away from real scenario. Nevertheless, the results obtained in \cite{rast08a} reveal that considering the sensors with high observation noise will cause severe decrease in performance of the distributed adaptive estimation algorithms such as IDLMS. To address this problem and to deal with the issue of reliability measurements, a new adaptive distributed estimation algorithm where each sensor participates in the algorithm according to its observation noise variance is proposed.

\subsection{Method}
To deal with the mentioned conditions, it is necessary to obtain an estimate of each sensor's observation noise. To do this, we consider the equation (1) again. If the IDLMS algorithm (i.e. (16)) is done for $Ls$ times (where $Ls$ is a suitably chosen integer), it is possible to have a primary estimate of  $w^o$. Now this primary estimate of $w^o$ is used to obtain each sensors observation noise. It must be noted that this estimate of $w^o$ is used just to obtain a primary estimate of observation noise at each sensor, and it is not the final estimate of $w^o$. Denoting by $\psi _k^{(Ls)} $ as the estimate of $w^o$ in the $i$th iteration in $N$th node we will have:

\begin{equation}
\psi _N^{(Ls)}  = \;\left. {\psi _k^{(i)} } \right|_{k = N,\,i = Ls} 
 \end{equation}

Using (1) and (17), the observation noise at each sensor can be estimated as

\begin{equation}
n_k (i) = d_k (i) - u_{k,i} \psi _N^{(Ls)} ,\quad i = 1,2,...,Ls
\end{equation}
 
In each node $k$, first the $n_k(i)$ is computed and then the variance of observation noise of the $k$th sensor is estimated by

\begin{equation}
\mathop {g_k }\limits_{k = 1,2,...,N}  = \left( {\frac{1}{{Ls}}} \right)\sum\limits_{i = 1}^{Ls} {n_k (i).\,} \,
\end{equation}

\begin{equation}
\tilde \sigma _k  = \sum\limits_{i = 1}^{Ls} {\left( {n_k (i) - g_k } \right)^2 }
\end{equation}

As $\tilde \sigma _k$ increases, the reliability of $d_k$ decreases, so there is inverse relation between $\tilde \sigma _k$ and sensor's reliability. Motivated by this fact we define the step-size of the our incremental distributed LMS algorithm as

\begin{equation}
\mu _{k}= \mu _{\max}e^{-a\tilde \sigma _k}
\end{equation}
where $\mu _{\max}$ is the global step-size parameter (which is constant for all sensors) and $a$ is a positive constant. It is obvious from the definition of (21) that larger observation noise variance (i.e. $\tilde \sigma _k$) yields smaller step-size parameter. Finally, for $i\geq Ls+1$ the IDLMS algorithm is modified as follows:
\begin{equation}
\psi _k^{(i)}  = \psi _{k - 1}^{(i)}  - \mu _k \left[ {R_{du,k}  - R_{u,k} w_{i - 1} } \right]^ * 
\end{equation}
After $i \to \infty$, all of the sensors will contain the appropriate estimate of $w^o$, that is 
\begin{equation}
\mathop {\lim }\limits_{i \to \infty } \;\psi _k^{(i)}  \to w^o,\ \;k\in \left\{1,2,\cdots,N\right\}  
\end{equation}

\section{Simulation Results}
In this section we present the simulation results of the proposed algorithm and compare it with the IDLSM algorithm of \cite{lop07}. To this aim, we consider a network with $N=30$ nodes and Gaussian regressors with $R_{u,k}=I$. We further assume that $\sigma _{v,k}^2 \in \left(10^{ - 3},10^{ - 1}\right)$. The curves are obtained by averaging over 100 experiments with $\mu _{\max}=0.01$ and $M=4$.
In Fig. 2, the performance of proposed algorithm for $Ls=20$ and $a=10$ in comparison with the IDLMS algorithm is depicted. To compare the performance of the mentioned algorithms we use the mean-square deviation (MSD) criteria which is defined as follows
\begin{equation}
{\rm{MSD}} \buildrel \Delta \over = E\left\|w^o-\psi_{k-1}^{(i)}\right\|^2.
\end{equation}

As it is clear from Fig. 2, the proposed algorithm has better performance in a sense of estimation performance. In Fig.3, the $\sigma ^{2}_{k}$ and the corresponding step-size parameter for each sensor in plotted.

\begin{figure}[t]
\centering
\includegraphics [width=8cm] {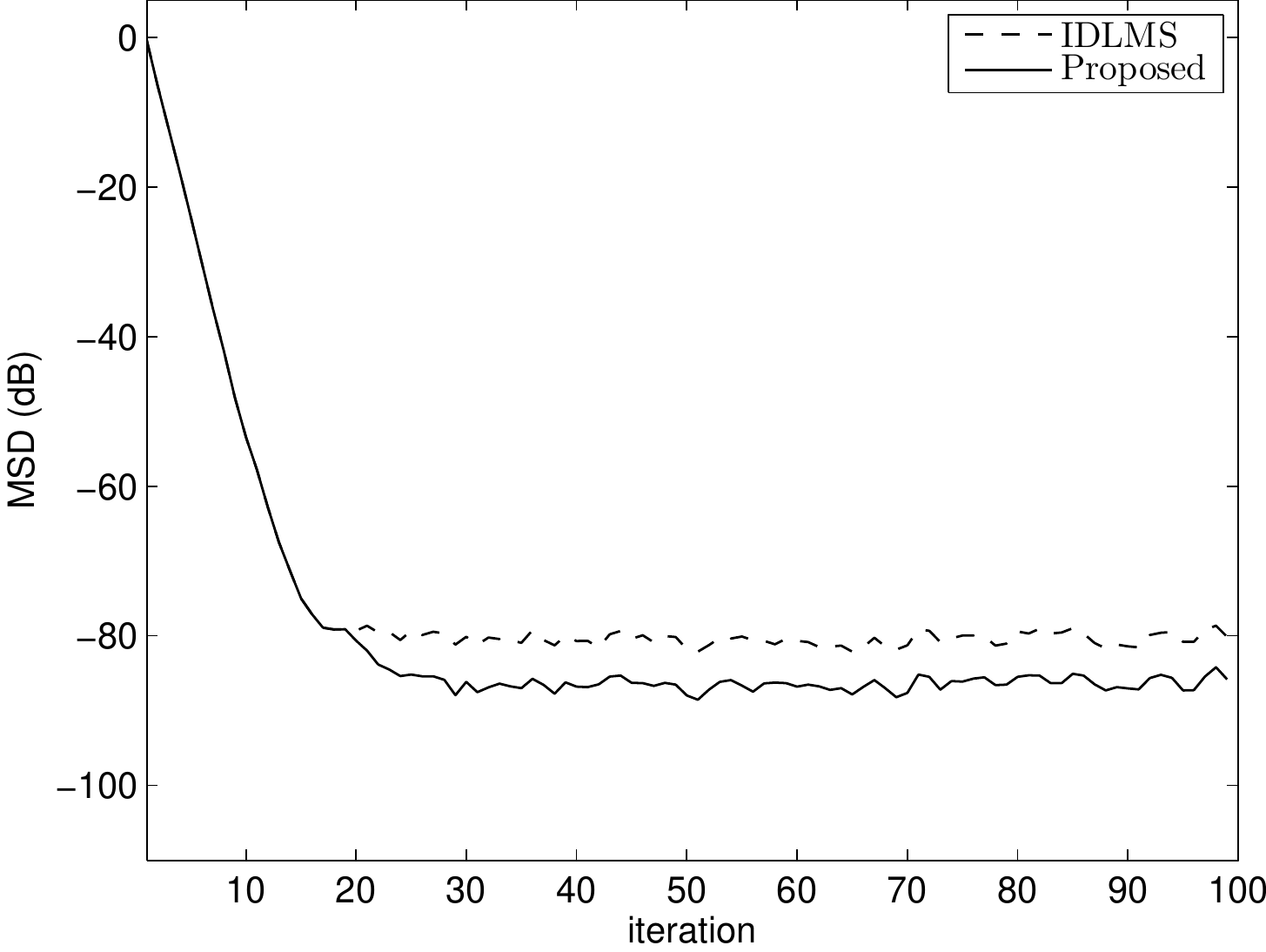}   
\caption{The MSD performance Proposed Algorithm with $Ls=10$ and $a$=10 in comparison with the IDLMS algorithm.}
\end{figure}

\begin{figure}[t]
\centering
\includegraphics [width=9cm] {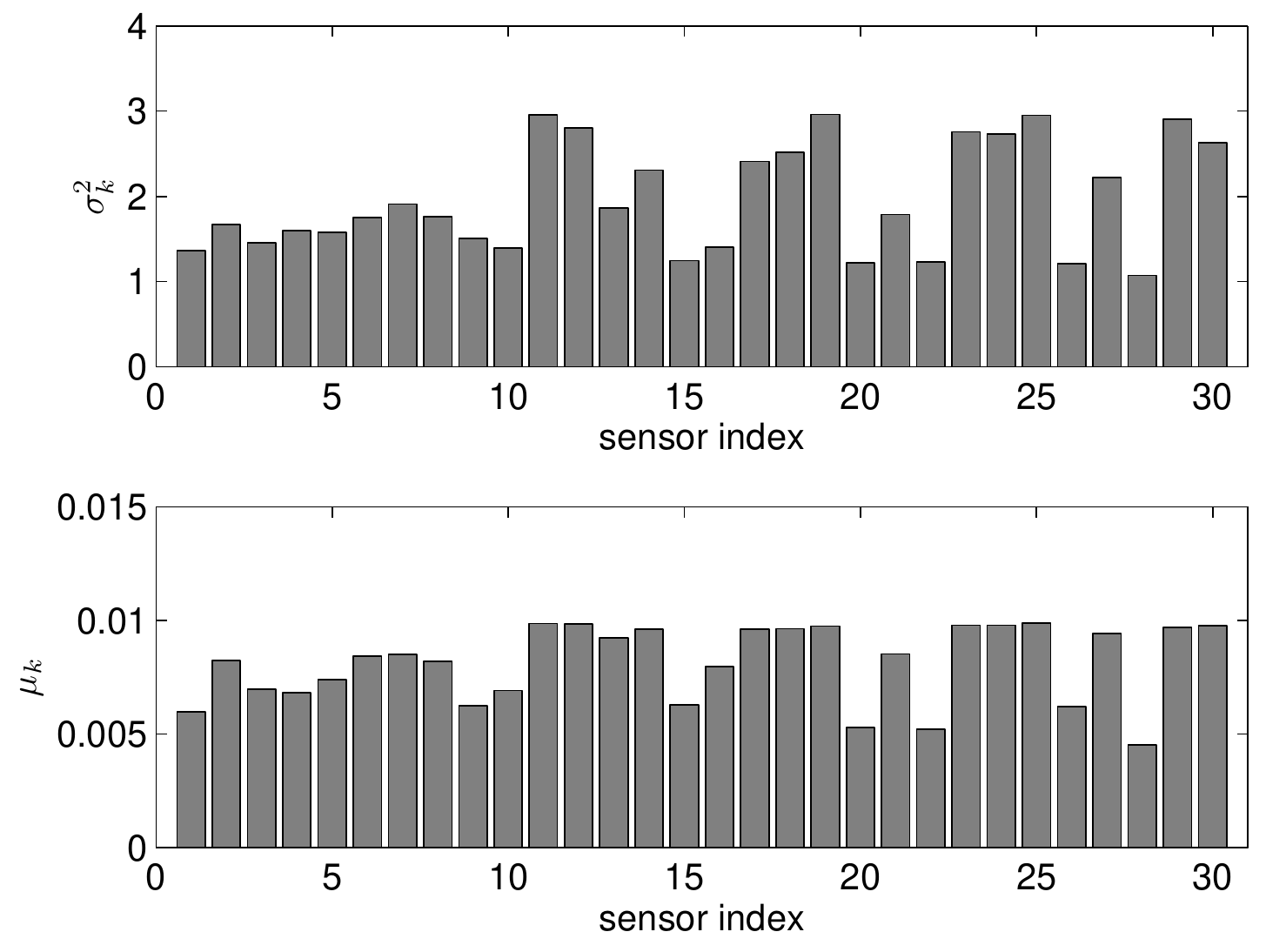}   
\caption{The $\sigma ^{2}_{k}$ (up) and the corresponding step-size parameter for each sensor (down).}
\end{figure}

The performance of the proposed algorithm depends on the value of $Ls$, since it determines how $\psi _k^{(Ls)}$  is close to $w^o$. In Fig. 4 the performance of the proposed algorithm for different values of $Ls$ in comparison with the IDLMS algorithm is shown. 
\begin{figure}[htbp]   
\centering
\includegraphics [width=9cm] {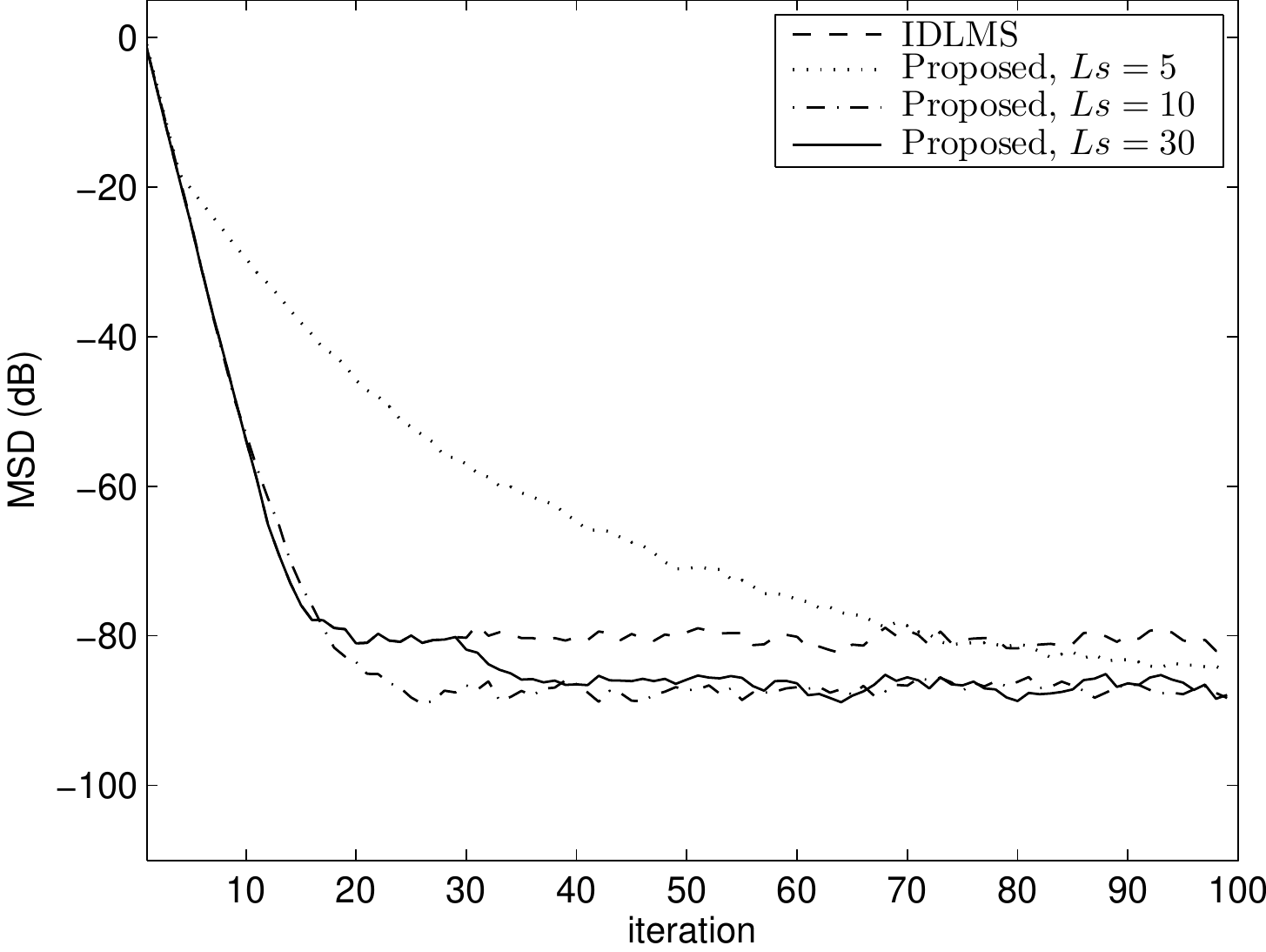}   
\caption{The performance of the proposed algorithm for $a=10$ and different values of $Ls$ in comparison with the IDLMS algorithm.}
\end{figure}
As it is clear from Fig. 4, as $Ls$ increases, better primary estimate of $w^o$ is obtained and as a result, a better final estimate of $w^o$ can be expected. It must be noticed that when the algorithm is in its steady-state, increasing the $Ls$ can not provide more better primary estimate of $w^o$. On the other hand, by choosing the $Ls$ such that the algorithm is not in its steady-state, the resulted $\psi _k^{(Ls)}$ is not close enough to $w^o$ which in turn makes a dramatically decrease in the performance of the proposed algorithm. These cases can be easily concluded from the Fig. 5 where the MSD performance of proposed algorithm for different values of $Ls$ is plotted. 

The performance of the proposed algorithm also depends on the $a$ parameter, (see (21)). By increasing $a$ ,the assigned step-size parameters become more smaller and as a result, the proposed algorithms provides better estimation performance (lower MSD) while, on the other hand, the convergence rate of proposed algorithm decreases. In Fig. 5 the performance of the proposed algorithm for different values of $a$ in comparison with the IDLMS algorithm is shown. 
\begin{figure}[htbp]    
\centering
\includegraphics [width=9cm] {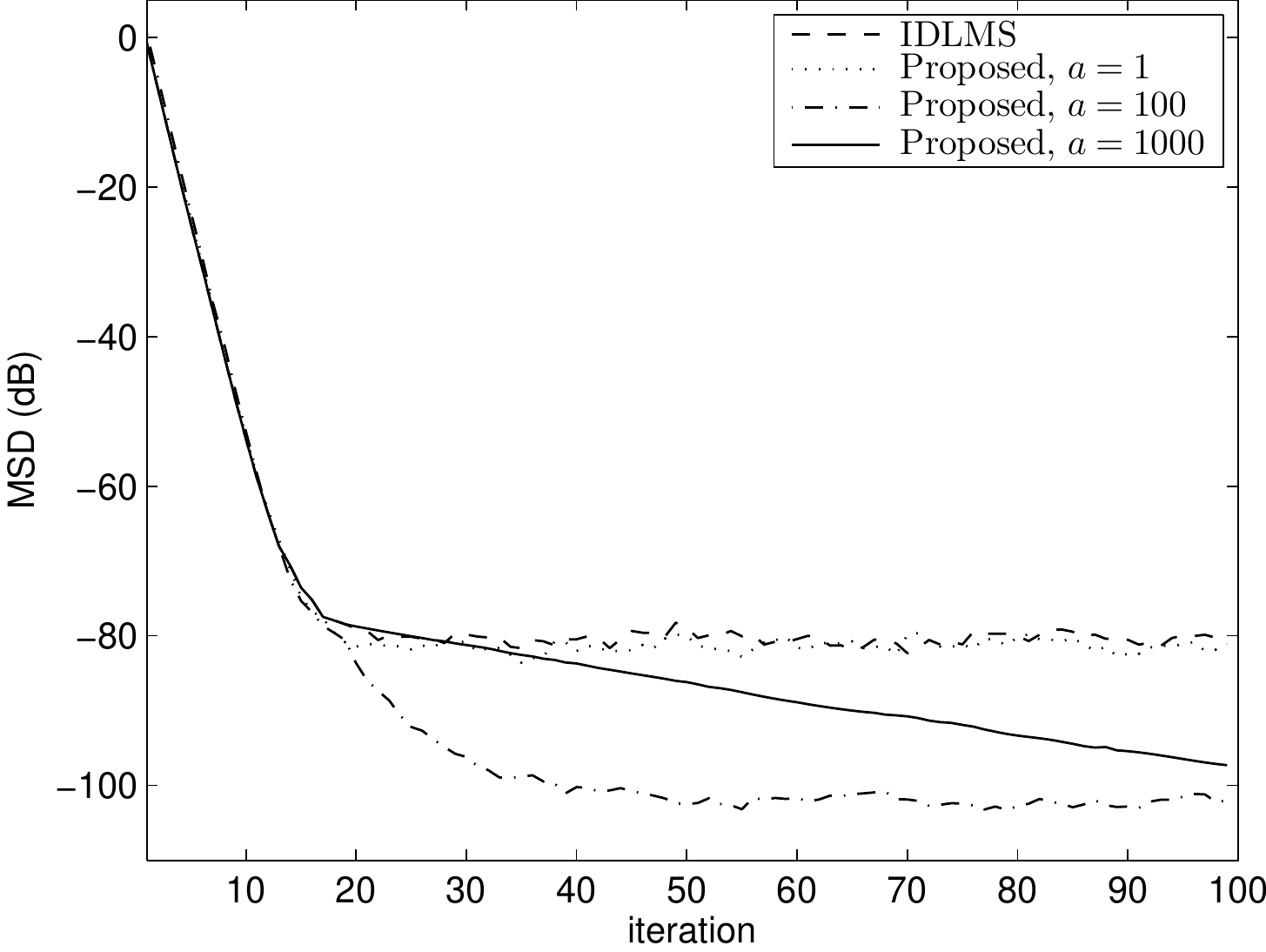}   
\caption{The performance of the proposed algorithm for $Ls=20$ and different values of $a$ in comparison with the IDLMS algorithm.}
\end{figure}
In the proposed algorithm by increasing the number of sensors in the network, the convergence rate of the algorithm decreases without change in the steady-state error which is the case for IDLMS algorithm. In Fig. 6 the performance of the proposed algorithm for different number of sensors,$K$ and $a=10$ and $Ls=20$ in comparison with the IDLMS algorithm is plotted.
\begin{figure}[htbp]   
\centering
\includegraphics [width=9cm] {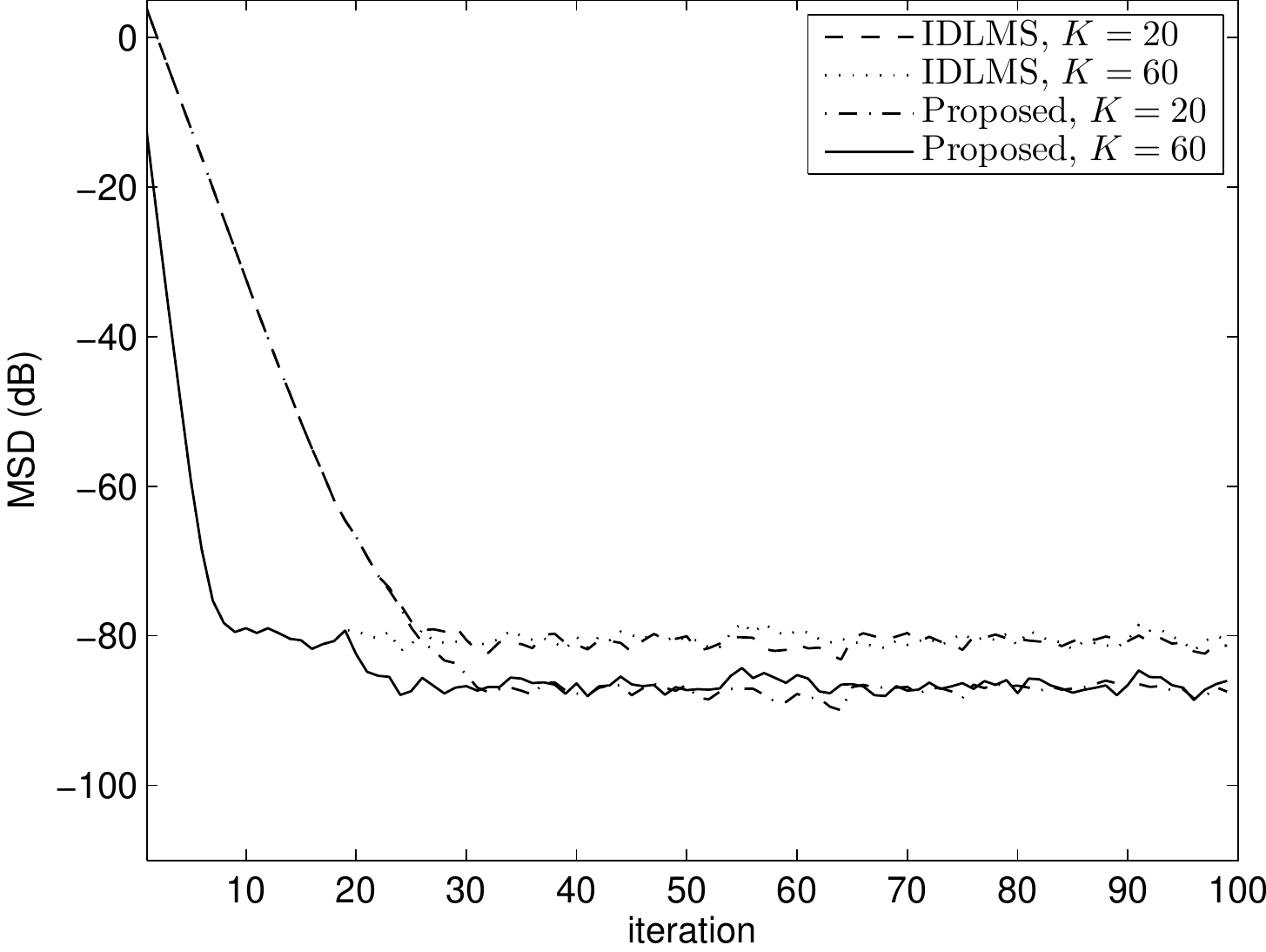}   
\caption{The performance of the proposed algorithm for $a=10$, $Ls=20$ and different number of sensors, i.e. $K$, in comparison with the IDLMS algorithm.}
\end{figure}
\section{Conclusion}
In this paper we considered the issue of reliability of measurements in distributed adaptive estimation algorithms. To deal with this issue we proposed a distributed adaptive estimation method based on IDLMS algorithm. The proposed algorithm contains two different phases: I) Estimating each sensor's observation noise and II) Estimating unknown parameter using the estimated observation noise variances. Also In this paper the step-size parameter is assigned to each sensor according to its observation noise variance. As the simulation results show, the proposed method outperforms the IDLMS algorithm in the sense of estimation error under the same conditions. It also must be noticed that although in this paper we consider the IDLMS algorithm as the base for our estimation method, the proposed method can be used in other adaptive estimation algorithm like diffusion least-mean square algorithm and DRLS as we did respectively in \cite{rast08b}.


%

%
%
%
%
%

\ifCLASSOPTIONcaptionsoff
  \newpage
\fi

\vfill

\end{document}